\documentstyle[aps]{revtex}
\begin{document}
\draft
\preprint{}
\title{Transport and conservation laws}
\author{X. Zotos$^1$, F. Naef$^1$ and P. Prelov\v sek$^{1,2}$}
\address{
$^1$Institut Romand de Recherche Num\'erique en Physique des
Mat\'eriaux (IRRMA), \\
INR-Ecublens, CH-1015 Lausanne, Switzerland}
\address{
$^2$J. Stefan Institute, University of Ljubljana, 6111 Ljubljana, Slovenia}
\bigskip\bigskip
\maketitle
\begin{abstract}
We study the lowest order conservation laws in one-dimensional (1D) 
integrable quantum many-body models (IQM) as the Heisenberg spin 1/2 chain, 
the Hubbard and t-J model. 
We show that the energy current is closely related to the first 
conservation law in these models and therefore the {\it thermal} transport 
coefficients are anomalous. Using an inequality on the time decay of current 
correlations we show how the existence of conserved quantities implies a 
finite charge stiffness (weight of the zero frequency component of the 
conductivity) and so ideal conductivity at finite temperatures.
\end{abstract}


Finite temperature transport properties, as electrical\cite{dr} and  
thermal\cite{kf} conductivity or spin dynamics\cite{ta} of 
strongly correlated 1D are recently in the center of experimental and 
theoretical interest. 
In this context, IQM's as the Heisenberg spin 1/2 chain, the Hubbard or $t-J$ 
model\cite{ba} are often the starting point for describing 1D systems.  
Further, from another point of view,  
the integrability of these models offers the possibility of studying  
the interplay between chaotic properties\cite{jl} and transport 
in quantum many-body systems, 
mostly studied in the context of classical many-body\cite{ford} or 
single particle quantum problems\cite{alt}.  

Based on analytical and numerical studies, we have 
recently proposed that IQM show anomalous, 
dissipationless finite temperature conductivity\cite{czp,zp,cz}. 
It is natural to think that this anomalous 
transport behavior is related to the macroscopic number of 
conserved quantities\cite{lu,sh,tet,gm} characterizing these systems.  
A set of conservation laws is represented by local involutive operators $Q_n$, 
commuting with each other $[Q_n,Q_m]=0$ and with the Hamiltonian, $[Q_n,H]=0$. 
The index $n$ indicates that the operator $Q_n$ is of 
the form $Q_n=\sum_{i=1}^L q_i^n$, where $q_i^n$ are local operators involving 
$n$ sites around site $i$, on a lattice of $L$ sites. 

Although rather formal procedures exist for the construction of these  
operators\cite{tet,gm}, it is not clear how to study their physical content 
and even more how to take them into account in the analysis of transport  
properties. 
In this paper we show, for different models of actual interest,  
that already the first 
nontrivial quantity $Q_3$ ($Q_2$ often denotes the Hamiltonian) has a simple 
physical significance: 
it is (or it is closely related to) the {\it energy current} operator. 
Further, we analyze how the coupling of the energy current or current 
operator to the conserved quantities results to time correlations not 
decaying to zero at long times. Thus, transport has not simple diffusive 
character and within the Kubo linear response theory\cite{kubo} 
is described by diverging or ill-defined transport coefficients.
 
We can relate the time decay of correlations to the local conserved 
quantities in the Hamiltonian systems we discuss by using an inequality 
proposed by Mazur\cite{m}:
\begin{equation}
\lim_{T\rightarrow \infty} \frac{1}{T} \int_0^T <A(t)A> dt \geq \sum_n 
\frac{<A Q_n>^2}{<Q_n^2>} ,
\label{mazur}
\end{equation}
where $< >$ denotes thermodynamic average,   
the sum is over a subset of conserved quantities ${Q_n}$, orthogonal 
to each other $<Q_n Q_m>=<Q_n^2>\delta_{n,m}$,   
$A^{\dagger}=A$ and $<A>=0$.  
In the following, we will only consider $Q_3$ in relation (\ref{mazur}) 
so that the issue of orthogonality will not enter. 
Further, we can write $<A(t)A>=C_{AA}+C(t)$ as the sum of a time-independent 
factor,
\begin{equation}
C_{AA}= 
\sum_a p_a \sum_{b (\epsilon_b=\epsilon_a)} |<a|A|b>|^2 
\label{maz2a}
\end{equation}
and a time-dependent one, 
\begin{equation}
C(t)= \sum_a p_a \sum_{b (\epsilon_b\neq \epsilon_a)} 
|<a|A|b>|^2 e^{i (\epsilon_a - \epsilon_b) t} .
\label{maz2b}
\end{equation}
Here $|a>, |b>$ are eigenstates of the Hamiltonian, 
$p_a=e^{-\beta \epsilon_a}/Z$ the corresponding Boltzmann weights and   
$\beta$ the inverse of the temperature.
For time correlations $<A(t)A>$ with non-singular low frequency 
behavior, the term $\lim_{T\rightarrow \infty} \frac{1}{T}
\int_0^T C(t) dt$ goes to zero 
and so $C_{AA}=\lim_{t\rightarrow \infty} <A(t)A>$,  
\begin{equation}
C_{AA} \geq \sum_n \frac{<A Q_n>^2}{<Q_n^2>} .
\label{maz3}
\end{equation}

In particular, we will use this inequality in the analysis of the real 
part of the conductivity,   
$\sigma'(\omega)=2\pi D(T)\delta(\omega)+\sigma_{reg}(\omega)$,  
related, within linear response theory, to the current-current 
correlation $<J(t)J>$. 
A finite value of the charge stiffness $D$, given also by 
$D=\frac{1}{2}\omega \sigma''(\omega)|_{\omega\rightarrow 0}$, implies an 
ideally conducting system\cite{kohn,czp}. 
We will now argue that $D \simeq \frac{\beta}{2L} C_{JJ}$ 
and therefore the following inequality holds for the charge stiffness:
\begin{equation}
D \geq (\frac{\beta}{2L}) \sum_n \frac{<J Q_n>^2}{<Q_n^2>} .
\label{eqd}
\end{equation}
In this derivation, we assume again that the regular part of the conductivity 
$\sigma_{reg}(\omega)$ shows a non-singular behavior at low frequencies so 
that the contribution from $C(t)$ in (\ref{mazur}) vanishes. This is a very 
mild condition for the physical systems we consider;  
actually, we even have indications\cite{zp,cz} from numerical simulations,  
that these IQM are characterized by a pseudogap and so a 
vanishing regular part $\sigma_{reg}(\omega\rightarrow 0)$. 

To relate $D$ to $C_{JJ}$, it is convenient to use a generalization of 
Kohn's approach\cite{kohn,czp} to finite temperatures. 
In this formulation $D(T)$ can be expressed as the thermal average of 
curvatures of energy levels in a Hamiltonian 
describing a system on a ring pierced by a fictitious flux $\phi$, 
$D=\frac{1}{2L} \sum_a p_a 
(\frac{\partial^2 \epsilon_a}{\partial\phi^2})|_{\phi\rightarrow 0}$. 
Evaluating the second derivative of the free energy $F$ as a function of  
the flux $\phi$ we find:
\begin{equation}
\frac{\partial^2 F}{\partial\phi^2}=
2LD-\beta\sum_a p_a (\frac{\partial\epsilon_a}{\partial \phi})^2
+\beta (\sum_a p_a \frac{\partial\epsilon_a}{\partial \phi})^2 .
\label{eqf}
\end{equation}
In the models we will discuss, the third term on the r.h.s. 
vanishes by symmetry (summing over $k$ and $-k$ subspaces). 
Further, these systems show no persistent 
currents at finite temperatures in the thermodynamic limit, therefore 
$\frac{\partial^2 F}{\partial\phi^2}|_{\phi\rightarrow 0}\rightarrow 0$ 
in this limit. We have numerically verified that this is indeed the case 
for temperatures larger than the level spacing;     
at zero temperature, there is no contradiction with 
Kohn's formula for $D$ as the second term in the r.h.s. vanishes in a 
ground state with zero current. 
Finally, as 
$\frac{\partial\epsilon_a}{\partial \phi}|_{\phi\rightarrow 0}=<a|j|a>$   
and degenerate levels contribute a vanishing weight in (\ref{maz2a}), 
we arrive at relation (\ref{eqd}). This inequality  
provides the lower bound for the charge stiffness $D$ which, if not zero,   
implies ideal conductivity at finite temperatures.

In general, it is difficult to evaluate the right hand side of the inequality 
(\ref{maz3}) involving the ``overlap" $<A Q_n>$. However,   
we will give some examples below, this correlation can easily be evaluated 
in the case of a grand canonical trace over states, 
in the thermodynamic limit and for $\beta \rightarrow 0$. We thus obtain the 
charge stiffness in leading order in $\beta$.

Before studying concrete models, we construct the energy current 
operator as follows: 
we consider Hamiltonians defined on a 1D lattice with $L$ sites 
and periodic boundary conditions $h_{L,L+1}=h_{L,1}$ of the form:
\begin{equation}
H=\sum_{i=1}^L h_{i,i+1} .
\label{H}
\end{equation}
Because the energy is a conserved quantity, the time evolution of the 
local energy operator $h_{i,i+1}(t)$ 
can be written as the discrete divergence of the energy current operator 
$J^E=\sum_{i=1}^L j^E_i$:
\begin{equation}
\frac{\partial h_{i,i+1}(t)}{\partial t}= i [H,h_{i,i+1}(t)]= 
- ( j^E_{i+1}(t) - j^E_i(t) ) ,
\label{je}
\end{equation}
where $h_{i,i+1}(t)=e^{i H t} h_{i,i+1} e^{-i H t}$ and 
$j^E_i= -i [h_{i-1,i},h_{i,i+1}]$. 
Now, by direct evaluation for several IQM, we will see that  
the energy current $J^E$ is closely related to a conserved quantity. 

\bigskip
\noindent
{\it i) Heisenberg model: } 
The general anisotropic Heisenberg Hamiltonian is given by:
\begin{equation}
H=\sum_{i=1}^L h_{i,i+1}=\sum_{i=1}^L (J_x S_i^x S_{i+1}^x + 
J_y S_i^y S_{i+1}^y + J_z S_i^z S_{i+1}^z) ,
\label{heis}
\end{equation}
where $S_i^{\alpha}=\frac{1}{2}\sigma_i^{\alpha}$,
$\sigma_i^{\alpha}$ are Pauli spin operators with components $\alpha=x,y,z$ 
at site $i$. 
The local energy current operator $j^E_i$ is:
\begin{equation}
j^E_i=J_x J_y (xzy - yzx)_{i-1,i+1} + J_y J_z (yxz - zxy)_{i-1,i+1} 
+ J_z J_x (zyx - xyz)_{i-1,i+1} ,
\label{jeheis}
\end{equation}
with $(\alpha\beta\gamma - \gamma\beta\alpha)_{i-1,i+1}=
S_{i-1}^{\alpha} S_i^{\beta} S_{i+1}^{\gamma} - 
S_{i-1}^{\gamma} S_i^{\beta} S_{i+1}^{\alpha}$. 
 
Now it is straigthforward to verify that the global energy current operator, 
$J^E=\sum_{i=1}^L j^E_i$, 
commutes with the Hamiltonian (\ref{heis}) and coincides with the first 
nontrivial conserved quantity, usually denoted by $Q_3$, obtained from an 
expansion of the transfer matrix in the algebraic Bethe ansatz 
method\cite{lu,gm}. 
In agreement with the notation $Q_2$ for the Hamiltonian,   
the local energy operator $h_{i,i+1}$ involves two sites $(i,i+1)$,  
while the local energy current operator $q^3_i=j^E_i$ involves three 
sites $(i-1,i,i+1)$.

The vanishing commutator $[J^E,H]=0$ implies that the 
energy current time correlations are independent of time.
\begin{equation}
<J^E(t) J^E>=\sum_a p_a j_{Ea}^2 ,
\label{kheis}
\end{equation}
where $j_{Ea}$ are the eigenvalues of $J^E$, $J^E|a>=j_{Ea}|a>$.   
The non-decaying of the energy current  
leads to a {\it diverging thermal conductivity} related to the 
$<J^E(t)J^E>$ correlation\cite{kubo}.

As for the conductivity, it is more relevant to discuss the fermionic 
version of the Heisenberg model, defined through the Jordan-Wigner 
transformation, the so called $t-V$ model:
\begin{equation}
H=(-t) \sum_{i=1}^L (c_{i}^{\dagger} c_{i+1} + h.c.)
+ V \sum_{i=1}^L (n_{i}-\frac{1}{2})(n_{i+1}-\frac{1}{2}) ,
\label{tv}
\end{equation}
where $c_{i}(c_{i}^{\dagger})$ denote annihilation (creation)
operators of spinless fermions at site $i$  
and $n_{i}=c_{i}^{\dagger}c_{i}$.

In this case, the corresponding energy current operator that commutes with   
the Hamiltonian (\ref{tv}) is given by:
\begin{equation}
J^E=\sum_i (-t)^2(i c_{i+1}^{\dagger} c_{i-1} + h.c.)+
V j_{i,i+1}(n_{i-1}+n_{i+2}-1) ,
\label{jqtv}
\end{equation}
where $j_{i,i+1}=(-t)(-i c_{i+1}^{\dagger} c_{i} + h.c.)$ is
the particle current.
Therefore, for this fermionic model, we find that the $<J^E(t) J^E>$  as well 
as the $<J^E(t) J>$ correlations are time independent implying 
a diverging thermal conductivity and ill-defined thermopower respectively.

Regarding the charge stiffness $D$, we can evaluate analytically   
$<J Q_3>^2/<Q_3^2>$  for $\beta \rightarrow 0$ and in the thermodynamic 
limit, obtaining from (\ref{eqd}), 
\begin{equation}
D\geq \frac{\beta}{2} \frac{2 V^2 \rho (1-\rho) (2\rho -1)^2} 
{1+V^2(2\rho^2-2\rho+1)} ,
\label{dtv}
\end{equation}
where $\rho$ is the fermion density. We note that for $\rho \neq 1/2$, 
$D$ is finite implying ideal conductivity as we have suggested 
before\cite{zp}. For $\rho=1/2$, this inequality is however insufficient 
for proving that $D$ is nonzero. 
Due to the electron - hole symmetry, this remains true 
even if we consider all the higher order 
conserved quantities $Q_n$. The reason is that, for the Heisenberg model,    
all $Q_n$'s can be 
generated\cite{tet,gm} by a recursive relation: $[B,Q_n]\sim Q_{n+1}$ where 
$B$ is a ``boost" operator given by: $B=\sum_n n h_{n,n+1}$. 
Then by the electron - hole transformation $c_i=(-1)^i \tilde c_i^{\dagger}$, 
we see that $J\rightarrow -J$ but $Q_n \rightarrow Q_n$ and  
therefore for $\rho=1/2$, $<J Q_n>=0$. The eventual nonorthogonality of the 
$Q_n's$ is not important as we can see by considering new orthogonal 
conserved quantities constructed using, for instance, a Gram-Schmidt 
orthogonalization procedure.

In Table \ref{table}, we present some indicative numerical results comparing 
$C_{JJ}$ with 
$<J Q_3>^2/<Q_3^2>$ for a couple of $\rho$ values and 
$\beta\rightarrow 0$.  
The results for $C_{JJ}$ were obtained by exact diagonalization of the 
Hamiltonian matrix on finite size lattices ($L$ up to 20 sites), 
followed by finite size scaling using a 2nd order polynomial in $1/L$. 
From this table we see that: 
i) the smaller the density the more the inequality (\ref{eqd}) is exhausted 
by just considering the contribution from $Q_3$, ii) 
for $V/t\rightarrow \infty$ the overlap $<J Q_3>$ gives the total 
weight of $C_{JJ}$; indeed, studying the 
higher order local conserved quantities we can see that they only contribute
terms in powers of $1/V$. Nevertheless, it is not clear why the 
inequality (\ref{maz3}) is exhausted and no other, e.g. nonlocal, conserved 
quantities contribute.

Finally, returning to the Heisenberg model, we note that the bound (\ref{dtv}) 
implies anomalous spin transport at finite magnetization, as the particle 
current maps to the spin current and the density $\rho$ to the magnetization  
(half-filling $\rho=1/2$ corresponding to zero magnetization).

\bigskip
\noindent
{\it ii) Hubbard model: }
It describes a system of interacting fermions on a lattice with 
Hamiltonian given by:

\begin{equation}
H=(-t) \sum_{\sigma,i=1}^L (c_{i\sigma}^{\dagger} c_{i+1 \sigma} + h.c.) 
+ U \sum_{i=1}^L (n_{i\uparrow}-\frac{1}{2})(n_{i\downarrow}-\frac{1}{2}) ,
\label{eq9}
\end{equation}
where $c_{i\sigma}(c_{i\sigma}^{\dagger})$ are annihilation (creation) 
operators of fermions with spin $\sigma=\uparrow, \downarrow$ at site $i$ 
and $n_{i\sigma}=c_{i\sigma}^{\dagger}c_{i\sigma}$.

Similarly as above, we can define a local energy operator by:
\begin{equation}
h_{i,i+1}=(-t)\sum_{\sigma} (c_{i\sigma}^{\dagger} c_{i+1 \sigma} + h.c.) 
+ \frac{U}{2} ( (n_{i\uparrow}-\frac{1}{2})(n_{i\downarrow}-\frac{1}{2})+
(n_{i+1\uparrow}-\frac{1}{2})(n_{i+1\downarrow}-\frac{1}{2}) ) .
\label{eq10}
\end{equation}
From the time evolution of $h_{i,i+1}$ we find the local energy current 
operator $j^E_i$ involving sites $(i-1,i,i+1)$:
\begin{equation}
j^E_i=\sum_{\sigma} (-t)^2(i c_{i+1\sigma}^{\dagger} c_{i-1\sigma} + h.c.)-
\frac{U}{2} (j_{i-1,i,\sigma}+j_{i,i+1,\sigma}) 
(n_{i,-\sigma}-\frac{1}{2}) ,
\label{eq11}
\end{equation}
where $j_{i,i+1\sigma}=(-t)(-i c_{i+1\sigma}^{\dagger} c_{i\sigma} + h.c.)$ is 
the particle current.
By comparing this expression for the energy current to the conserved 
quantity\cite{sh,gm} $Q_3$, we find that they coincide when the factor $U/2$ 
in (\ref{eq11}) is replaced by $U$. 
So the energy current $J^E=\sum_{i=1}^L j_i^E$ does not commute with the 
Hamiltonian.   
However, as $J^E$ has a finite overlap $<J^E Q_3>$, with the conserved
quantity $Q_3$ we still find that the energy
current correlations decay to a finite value at long times   
so that the thermal transport coefficients are anomalous. We can find a 
lower bound for the decay by using (\ref{mazur})  
for $\beta\rightarrow 0$ and in the thermodynamic limit:
\begin{equation}
\lim_{t\rightarrow\infty} <J^E(t)J^E>=C_{J^E J^E}  \geq
\frac{<J^E Q_3>^2}{<Q_3^2>}
\label{klim}
\end{equation}
\begin{equation}
\frac{<J^E Q_3>^2}{<Q_3^2>}=L \sum_{\sigma} 2\rho_{\sigma}(1-\rho_{\sigma})+
\frac{U^4}{4} \frac
{[\sum_{\sigma}2\rho_{\sigma}(1-\rho_{\sigma})
(2\rho_{-\sigma}^2-2\rho{-\sigma}+1)]^2}
{\sum_{\sigma} 2\rho_{\sigma}(1-\rho_{\sigma})[1+U^2(2\rho_{-\sigma}^2-
2\rho_{-\sigma}+1)]} .
\label{kappa}
\end{equation}

As for the charge stiffness $D$, we can again evaluate analytically
$<J Q_3>^2/<Q_3^2>$  for $\beta \rightarrow 0$ and in the thermodynamic
limit, obtaining from (\ref{eqd}),
\begin{equation}
D \geq \frac{\beta}{2} \frac
{[U\sum_{\sigma}2\rho_{\sigma}(1-\rho_{\sigma})(2\rho_{-\sigma}-1)]^2}
{\sum_{\sigma} 2\rho_{\sigma}(1-\rho_{\sigma})[1+U^2(2\rho_{-\sigma}^2-
2\rho_{-\sigma}+1)]} ,
\label{dhubb}
\end{equation}
where $\rho_{\sigma}$ are the densities of $\sigma=\uparrow,\downarrow$ 
fermions. 
For $\rho_{\sigma}=1/2$, the right hand side of (\ref{dhubb}) 
vanishes, although a general proof involving all higher conserved quantities 
is not possible as a boost operator for the Hubbard model is not known.

\bigskip
\noindent
{\it iii) ``t-J" model: }
It belongs to a class of multicomponent quantum systems\cite{suth} 
describing interacting particles of different species, 
singly occupying each site.
The Hamiltonian acts on each bond $(i,i+1)$ by the operator $P_{i,i+1}$ which 
permutes neighboring particles, independently of their type:
\begin{equation}
H=\sum_{i=1}^L P_{i,i+1}
\label{perm}
\end{equation}
For this generic model we can directly verify that the energy current operator
\begin{equation} 
J^E=-i \sum_{i=1}^L [P_{i-1,i},P_{i,i+1}] , 
\label{pje}
\end{equation}
coincides with a conserved quantity\cite{kor} and so commutes with 
the Hamiltonian.

Now considering three types of particles, corresponding to empty sites, 
up spins and down spins, we recover the $t-J$ model\cite{kor} for special 
values of $J/t$. 
This model describes a system of interacting fermions subject to a constraint 
of no double occupancy, with Hamiltonian given by:
\begin{equation}
H=-t \sum_{\sigma,i=1}^L P (c_{i \sigma}^{\dagger} 
c_{i+1 \sigma} + h.c.)P 
+ J \sum_{i=1}^L (\vec S_i \vec S_{i+1}-n_i n_{i+1}/4)+2\hat N-L ,
\label{eqtj}
\end{equation}
where $c_{i\sigma} (c_{i\sigma}^{\dagger})$ are annihilation 
(creation) operators of a fermion on site $i$ with spin 
$\sigma=\uparrow, \downarrow$. 
$P=\prod_{i=1}^{L} (1-n_{i\uparrow}n_{i\downarrow})$ 
is a projection operator on sites with no double occupancy,
$n_{i\sigma}=c_{i\sigma}^{\dagger}c_{i\sigma}$, 
$\hat N=\sum_{i=1,\sigma}^L n_{i\sigma}$. 

This model is integrable for $J/t=0$, corresponding to the 
$U/t\rightarrow \infty$ limit of the Hubbard model or to the 
model (\ref{perm}) where permutations act only on bonds with 
``empty - up spin" or ``empty - down" configurations.
For this case, we found that the corresponding energy current commutes 
with the Hamiltonian, as is also known for the particle current\cite{rice}. 
Finally, for $J=2t$, the ``supersymmetric" model (\ref{eqtj}) is also 
integrable and the 
energy  current coincides with the conserved quantity $Q_3$ as is 
presented in reference (19). Therefore the transport coefficients 
of the supersymmetric $t-J$ model are also anomalous. 

\bigskip
The above results imply that, at least, certain quantities 
related to transport coefficients in IQM are non-ergodic 
(see however reference \cite{jl} for the recent notion of ``mixing"). 
Within linear response theory, 
this translates to ideal conducting behavior at finite temperatures, 
the charge stiffness $D$ being a measure of nonergodicity. We also expect 
that noise, as is described by the spectral properties of the current-current 
correlations, shows anomalous behavior, more characteristic of a 
ballistic rather than a diffusive system. 
This behavior is to be contrasted to the 
normal dissipative behavior we found\cite{zp,cz} for similar   
nonintegrable systems with no conservation laws.
So IQM, on the one hand are not generic models for studying 
finite temperature transport in quantum many-body systems,  
but on the other hand they offer the possibility of observing new effects.  
The main remaining question of course, is the robustness of (nearly) ideal 
conducting behavior for systems close to integrability,   
a problem very similar to the one in classical near-integrable 
nonlinear systems. Anyway, 
as it is known that conservation laws play an important role on transport 
properties\cite{forster} they should be taken into account in approximate 
or exact analysis of these properties.

\acknowledgments
We would like to thank F. D. M. Haldane, H. Kunz, T. M. Rice and 
C. Stafford for useful discussions.  
This work was supported by the Swiss National Fond Grants No. 20-39528.93,
the University of Fribourg, the Ministry of Science and Technology of 
Slovenia and the Institute of Scientific Interchange.

\begin{table}
\caption{$\frac{<J Q_3>^2}{<Q_3^2>}/C_{JJ}$ as a function of 
filling $\rho$ and $V/t$.}
\begin{tabular}{c|cc}
$V/t$ & $\rho=1/3$ & $\rho=1/4$ \\
\tableline
0.0 & 0.0 & 0.0 \\
1.0 & 0.11 & 0.23 \\
2.0 & 0.50 & 0.58 \\
4.0 & 0.83 & 0.89 \\
8.0 & 0.96 & 0.98 \\
$\infty$ & 1.0 & 1.0 \\
\end{tabular}
\label{table}
\end{table}

\begin{references}
\bibitem{dr} M. Dressel, A. Schwartz, G. Gr\"uner and L. Degiorgi, 
Phys. Rev. Lett. {\bf 77}, 398 (1996).
\bibitem{kf} C. L. Kane and M. P. A. Fisher, Phys. Rev. Lett. {\bf 76}, 
3192 (1996) and references therein.
\bibitem{ta} M. Takigawa, N. Motoyama, H. Eisaki and S. Uchida, Phys. Rev. 
Lett. {\bf 76}, 4612 (1996); P.H.M. van Loosdrecht {\it et al.}, Phys. 
Rev Lett. {\bf 76}, 311 (1996).
\bibitem{ba} V. E. Korepin, N.M. Bogoliubov and A. G. Izergin, 
{\it Quantum inverse scattering method and correlation functions} 
(Cambridge University Press, 1993).
\bibitem{jl} G. J. Lasinio and C. Presilla, Phys. Rev. Lett. {\bf 77}, 
November 1996.
\bibitem{ford} J. Ford, Phys. Rep. {\bf 213}, 271 (1992).
\bibitem{alt} N. Taniguchi and B. L. Altshuler, Phys. Rev. Lett. {\bf 71}, 
4031 (1993).
\bibitem{czp} H. Castella, X. Zotos, P. Prelov\v sek, Phys. Rev. Lett. 
{\bf 74}, 972 (1995).
\bibitem{zp} X. Zotos and P. Prelov\v sek, Phys. Rev. B{\bf 53}, 983 (1996).
\bibitem{cz} H. Castella and X. Zotos, Phys. Rev. B{\bf 54}, 4375 (1996).
\bibitem{lu} M. L\"uscher, Nucl. Phys. B{\bf 117}, 475 (1976).
\bibitem{sh} B.S. Shastry, Phys. Rev. Lett. {\bf 56}, 1529 (1986);  
E. Olmedilla and M. Wadati, Phys. Rev. Lett. {\bf 60}, 1595 (1988).
\bibitem{tet} M. G. Tetel'man, Sov. Phys. JETP {\bf 55}, 306 (1982);  
H. B. Thacker, Physica D{\bf 18}, 348 (1986).
\bibitem{gm} M. P. Grabowski and P. Mathieu, Ann. Phys. {\bf 243}, 299 (1996).  
\bibitem{kubo} R. Kubo, J. Phys. Soc. (Japan), {\bf 12}, 570 (1957).
\bibitem{m} P. Mazur, Physica {\bf 43}, 533 (1969); 
M. Suzuki, Physica {\bf 51}, 277 (1971).
\bibitem{kohn} W. Kohn, Phys. Rev. {\bf 133}, 171 (1964). 
\bibitem{suth} B. Sutherland, Phys. Rev. B{\bf 12}, 3795 (1975);  
P. P. Kulish and N. Yu. Reshetikhin, JETP {\bf 53}, 108 (1981).
\bibitem{kor} F. H. L. Essler and V. E. Korepin, 
Phys. Rev. B{\bf 46}, 9147 (1992).
\bibitem{rice} W. F. Brinkman and T. M. Rice, Phys. Rev. B{\bf 2}, 1324 (1970).
\bibitem{forster} D. Forster, {\it Hydrodynamic fluctuations,  
broken symmetry and correlation functions}, (W. A. Benjamin, New York, 1975).

\end{references}
\end{document}